\RequirePackage{fix-cm}
\documentclass[twocolumn]{svjour3}          % twocolumn
\smartqed  % flush right qed marks, e.g. at end of proof
\usepackage{natbib}
\usepackage{multirow}
\usepackage{graphicx}
\usepackage{amsmath}
\usepackage{xcolor}

\usepackage{ulem} % use normalem to protect \emph
\newcommand{\hl}[1]{#1}

\newcommand{\hz}[1]{#1}

\makeatletter
\renewcommand*\env@matrix[1][\arraystretch]{%
  \edef\arraystretch{#1}%
  \hskip -\arraycolsep
  \let\@ifnextchar\new@ifnextchar
  \array{*\c@MaxMatrixCols c}}
\makeatother

\makeatletter
\newcommand\tabfill[1]{%
  \dimen@\linewidth
  \advance\dimen@\@totalleftmargin
  \advance\dimen@-\dimen\@curtab
  \parbox[t]\dimen@{#1\ifhmode\strut\fi}%
}
\makeatother

\journalname{Rock Mechanics and Rock Engineering - Accepted Manuscript}
\begin{document}

\title{Thermo-poro-elastic behaviour of a transversely isotropic shale: Thermal expansion and pressurization
%\title{Transversely isotropic thermo-poro-elastic behaviour of the Callovo-Oxfordian claystone: Thermal pressurization and expansion}
%\title{Thermal behaviour of the Callovo-Oxfordian claystone: Transversely isotropic deformations and thermal pressurization 
}

\titlerunning{Thermal behaviour of the Callovo-Oxfordian claystone}        % if too long for running head

\author{Philipp Braun         \and
        Siavash Ghabezloo     \and
        Pierre Delage         \and
        Jean Sulem        \and
        Nathalie Conil       
}

%\authorrunning{Short form of author list} % if too long for running head

\institute{           P. Braun         \and
        	S. Ghabezloo     \and
        	P. Delage         \and
        	J. Sulem 		\at
              Laboratoire Navier, 6-8 avenue Blaise-Pascal, Cité Descartes
77455 Champs-sur-Marne, Paris, France, philipp.braun@enpc.fr
           \and
              N. Conil        
			\at
              Andra, Meuse/Haute-Marne Underground Research Laboratory, Bure, France \\
  }

\date{Accepted: 5 October 2020}
% The correct dates will be entered by the editor

\onecolumn

\maketitle

\begin{abstract}
The Callovo-Oxfordian (COx) claystone is considered as a candidate host rock for a deep geological radioactive waste repository in France. Due to the exothermic waste packages, the rock is expected to be submitted to temperatures up to 90 $^{\circ}$C. The temperature rise induces deformations of the \hl{host rock}, together with an increase in pore pressures, involving complex thermo-hydro-mechanical (THM) couplings. This study aims to better characterize the THM response of the COx claystone \hl{to temperature changes} in the laboratory. To this end, claystone \hl{specimens} were tested in a temperature controlled, high pressure isotropic \hl{compression} cell, under stress conditions close to the in-situ ones. Thermal loads were applied on the specimens along different heating and cooling paths. A temperature corrected strain gage system provided precise measurements of the anisotropic strain response of the specimens. Drained and undrained thermal expansion coefficients in both transversely isotropic directions were determined. The measurement of pore pressure changes in undrained condition yielded the thermal pressurization coefficient. All parameters were analysed for their compatibility within the thermo-poro-elastic framework, and their stress and temperature dependency was identified.  

% \PACS{PACS code1 \and PACS code2 \and more}
% \subclass{MSC code1 \and MSC code2 \and more}
\end{abstract}

\clearpage
\twocolumn
\section*{List of Symbols}

\hl{Note that the matrix notation is used throughout this work.}
\begin{tabbing}
\hspace{0.1\columnwidth}  \= \hspace{0.1\columnwidth}   \= \hspace{0.8\columnwidth}    \kill \ignorespaces
$\varepsilon_i$ \> \tabfill{ Strain \hl{vector containing the 6 independent components of the second rank strain tensor}}\\
$\varepsilon_v$ \> \tabfill{ Volumetric strain }\\
$C_{ij}$ \> \tabfill{ Drained compliance tensor in matrix format	}\\
$\sigma_i$ \> \tabfill{ Stress \hl{vector containing the 6 independent components of the second rank stress tensor}}\\
$\sigma$ \> \tabfill{ Isotropic confining stress	}\\
$\sigma'$ \> \tabfill{ Terzaghi \hl{isotropic} effective stress}\\
$b_i$ \> \tabfill{ Biot's coefficient for $i$-th direction	}\\
$p_f$ \> \tabfill{ Pore fluid pressure	}\\
$T$ \> \tabfill{ Temperature	}\\
$\alpha_{d,i}$ \> \tabfill{ Linear drained thermal expansion coefficient in the $i$-th direction}\\
$\alpha_{d}$ \> \tabfill{ Volumetric drained thermal expansion coefficient}\\
$H_{i}$ \> \tabfill{ Biot's \hl{linear pore pressure loading} modulus in the $i$-th direction			}\\
$H$ \> \tabfill{ Biot's \hl{volumetric pore pressure loading} modulus}\\
$K_{d}$ \> \tabfill{ \hl{Drained} bulk modulus			}\\
$K_{s}$ \> \tabfill{ \hl{Unjacketed} bulk modulus}\\
$\alpha_{u,i}$ \> \tabfill{ Linear undrained thermal expansion coefficient in the $i$-th direction		}\\
$\alpha_{u}$ \> \tabfill{ Volumetric undrained thermal expansion coefficient		}\\
$\Lambda$ \> \tabfill{ Thermal pressurization coefficient		}\\
$\alpha_{f}$ \> \tabfill{ Volumetric thermal expansion coefficient of the pore fluid		}\\
$\alpha_{\phi}$ \> \tabfill{ Volumetric thermal expansion coefficient of the pore space		}\\
${\phi}$ \> \tabfill{ Porosity		}\\
$K_{\phi}$ \> \tabfill{ \hl{Unjacketed pore} modulus				}\\
$K_{f}$ \> \tabfill{ Bulk modulus of the pore fluid				}\\
$\Lambda^{\rm{mes}}$ \> \tabfill{ Measured thermal pressurization coefficient		}\\
$\Lambda^{\rm{cor}}$ \> \tabfill{ Corrected thermal pressurization coefficient		}\\
$\Lambda_L$ \> \tabfill{ Thermal pressurization coefficient of the drainage system		}\\
$V_L$ \> \tabfill{ Volume of the drainage system		}\\
$c_L$ \> \tabfill{ Compressibility of the drainage system		}\\
$c_f$ \> \tabfill{ Compressibility of the pore fluid		}\\
$V$ \> \tabfill{ Specimen volume		}\\
$\alpha^{\rm{mes}}_{u}$ \> \tabfill{ Measured volumetric undrained thermal expansion coefficient		}\\
$\alpha^{\rm{cor}}_{u}$ \> \tabfill{ Corrected volumetric undrained thermal expansion coefficient		}\\
$\alpha_{d,i}^*$ \> \tabfill{ Elasto-plastic drained thermal expansion coefficient in the $i$-th direction}\\
$\alpha_{d,i}^{\rm{irr}}$ \> \tabfill{ \hl{Inelastic} drained thermal expansion coefficient in the $i$-th direction}\\
$\eta$ \> \tabfill{ Ratio between pore water and bulk water thermal expansion coefficients}\\
\hl{$\kappa$} \> \tabfill{ \hl{Model parameter for the temperature dependency of $H$}}\\
$\rho$ \> \tabfill{ Wet density}\\
$\rho_d$ \> \tabfill{ Dry density}\\
$w$ \> \tabfill{ Water content}\\
$S_r$ \> \tabfill{ Saturation degree}\\
$s$ \> \tabfill{ Suction}\\
$\varepsilon_{\mathrm{hyd}}$ \> \tabfill{ \hl{Hydration swelling} }\\
\end{tabbing}

\section{Introduction}
\label{sec:thermal:1}
The identification of thermo-poro-elastic parameters of host rocks for deep geological radioactive waste disposal is of great importance to predict the effect of the heat generated from exothermic waste packages. The microtunnels, in which the exothermic waste packages will be placed, are designed \hl{with sufficient distance to each other, as} to prevent temperatures larger than 90 $^{\circ}$C in the host rock. Various in-situ and laboratory studies have been carried out \hl{among others} on the COx claystone, as well as on the Opalinus claystone, a possible host rock in Switzerland with comparable characteristics. 

\hl{In undrained conditions, a rise of temperature in claystone does not only generate thermal strains, but can also induce an increase in pore pressure, called thermal pressurization \mbox{\citep{Ghabezloo200901}}.} \hl{This phenomenon has been measured in-situ, during the TER test in COx claystone in the Bure underground laboratory and during the HE-D test in Opalinus clay in the Mont Terri rock laboratory (see for instance \mbox{\citealp{Gens200720}} and \mbox{\citealp{Seyedi201775}}). 
Also in the laboratory, thermal pressurization, characterized through the thermal pressurization coefficient, has been investigated on COx samples \mbox{\citep{Mohajerani201211,Zhang201746}}. These authors indicated, that the thermal pressurization is due to different coupled THM properties, which can vary with stress and/or temperature. Hence also the thermal pressurization coefficient can be dependent on both temperature and stress conditions. }

\hl{In particular, thermal pressurization of claystones is governed \hz{by} the thermal expansion and elastic properties of both rock matrix and pore fluid. \mbox{\cite{Gens200720}} concluded in their numerical sensitivity analysis of in-situ tests, that the most important coupling occurred between thermal and hydraulic behaviour, when the relatively high thermal expansion coefficient of water generates pore pressure. This is followed by the coupling between hydraulic and mechanical properties, as the increasing pore pressure causes deformation of the rock matrix. Here one also has to take into account the anisotropic nature of stiff clays. Considering the importance of the thermal expansion properties of solid and fluid phases, these authors found that a wide range of values from laboratory experiments exist in this regard.}

\hl{Different studies aimed therefore to identify separately these THM characteristics and back-calculate the thermal pressurization coefficient, using a thermo-poro-elasticity framework \mbox{\citep{Ghabezloo200901}}:}
\hl{In drained heating tests, the thermal expansion properties of the rock matrix were investigated. \mbox{\cite{Monfared201173}} observed the thermal volume changes of Opalinus clay, where under heating up to 65 $^{\circ}$C the specimens expanded, prior to exhibiting contraction at higher temperatures. This contraction was of plastic nature, evidenced after a subsequent cooling and re-heating cycle. \mbox{\cite{Belmokhtar201722}} made similar observations above 48 $^{\circ}$C on the COx claystone. The temperature, at which plastic contraction occurred, was suspected to be close to the maximum temperature to which the material was previously exposed to during its geological history (around 50 $^{\circ}$C for COx claystone according to \mbox{\citealp{Blaise201471}}). Such thermoplastic behaviour has also been observed on clay, for which thermal expansion properties can depend on its degree of consolidation. Various clays exhibit a thermoelastic expansion when in overconsolidated state, and thermoplastic contraction in normally consolidated state (e.g. \mbox{\citealp{Baldi198880,Sultan200213}} on Boom clay, \mbox{\citealp{Abuel200714}} on Bangkok clay).}

\hl{Moreover, peculiar features of the thermal expansion coefficient of the pore water in clayey geomaterials have been indicated. For instance, \mbox{\cite{Monfared201173}} presented some data on the undrained thermal behaviour of Opalinus clay, that could only be reproduced by adopting an anomalously high thermal expansion coefficient of the pore fluid. Similar observations have been previously made by \mbox{\cite{Baldi198880}} on low porosity clays. Indeed, \mbox{\cite{Derjaguin199239}} and \mbox{\cite{Xu200950}} reported this anomaly on water contained in small pores of silica gel and silica glasses, respectively. This phenomenon has hitherto not been observed on the COx claystone.}

\hl{In our companion paper \mbox{\citep{Braun2020a}}, we identified the hydro-mechanical properties of the COx claystone in isothermal conditions. Laboratory experiments were conducted in drained and undrained state, applying total stress changes and hydraulic loading. This provided elastic coefficients, which constitute the transversely isotropic stiffness matrix and the transversely isotropic Biot coefficients of the material. These coefficients, taking into account their change with Terzaghi isotropic effective stress, were used in the present work for the back-analysis of the thermal pressurization coefficient.}

\hl{The experimental study presented in this work was carried out to further characterize the thermal behaviour of the COx claystone. Complementing the data provided by \mbox{\cite{Mohajerani201211}}, \mbox{\cite{Mohajerani201413}} and \mbox{\cite{Belmokhtar201722}}, we aim for obtaining a complete set of transversely isotropic THM parameters, able to describe strains and pore pressure changes occurring under thermal loads. COx samples were tested along various heating and cooling paths under constant total stress. This was done in an isotropic high pressure loading cell \mbox{\citep{Tang200845}}, improved by a temperature corrected strain gage system. The device permitted precise deformation measurements on the specimens, allowing the determination of drained and undrained thermal expansion coefficients, as well as of the thermal pressurization coefficient.  }

\section{Thermo-Poro-Elastic Framework}
\label{sec:thermal:2}

%\subsection{Transverse isotropy}
%\label{sec:thermal:3.1}

The COx claystone is trans\-verse\-ly i\-so\-trop\-ic, with \hl{lower} stiffness and \hl{higher} thermal expansion coefficients perpendicular to the bedding plane (\mbox{\citealp{Chiarelli2000}}; \mbox{\citealp{Escoffier2002}}; \mbox{\citealp{Andra2005}}; \mbox{\citealp{Mohajerani201211}}; \mbox{\citealp{Zhang201279}}; \mbox{\citealp{Belmokhtar201722,Belmokhtar201787}}). 

\begin{figure}[htbp]
% Use the relevant command to insert your figure file.
% For example, with the graphicx package use
  \includegraphics[width=0.85\columnwidth]{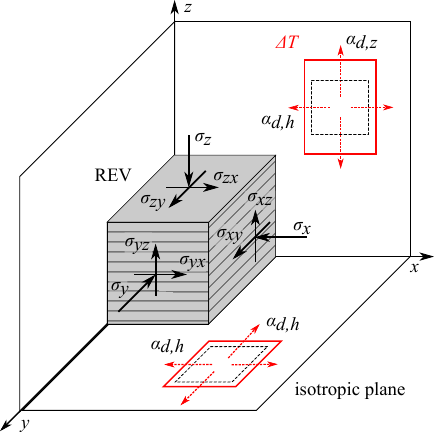}%{cell.pdf}
% figure caption is below the figure
\caption{\hl{Representative elementary volume (REV) in a transversely isotropic frame, adopted from \mbox{\cite{Popov201920}}. Transversely isotropic layering is indicated on the REV, resulting in a isotropic plane $x=y=h$ with the direction $z$ perpendicular to the layers. Upon temperature change $\Delta T$, the REV deforms according to $\alpha_{d,i}$. }}
\label{fig:thermal:trans_iso_T}       % Give a unique label
\end{figure}

The thermal properties and THM couplings in the COx claystone are analysed within a thermo-poro-elastic constitutive framework, reviewed amongst others by \cite{Biot195759}, \cite{Palciauskas198228} and \cite{Coussy2004}. \hl{In this work, we use the matrix notation, \hz{with} stress and strain vectors $\sigma_i$ and $\varepsilon_{i}$, containing the 6 independent components of their corresponding second rank tensors.} Following the formulation of \cite{Cheng199719} for a transversely isotropic material, we express a change of strains $\varepsilon_{i}$ with respect to a change of stresses $\sigma_i$, pore fluid pressure $p_f$ and temperature $T$ as a sum of partial derivatives: 
\begin{equation}
\label{eq:thermal:epsi}
{\rm{d}}{\varepsilon _i}{\rm{ = }}{C_{ij}}{\rm{d}}{\sigma _j} - {C_{ij}}{b_j}{\rm{d}}p_f - {\alpha_{d,i}}{\rm{d}}T
\end{equation}
\hl{where $C_{ij}$ is the transversely isotropic compliance matrix, \hz{$b_{j}$} a vector of transversely isotropic Biot's coeffients, described by \mbox{\cite{Cheng199719}}, and $\alpha_{d,i}$ a vector of linear drained thermal expansion coefficients.} 
\hl{The spatial orientation\hz{s} of the stress/strain tensor and the thermal expansion properties \hz{are} depicted on a elementary volume in Fig. \ref{fig:thermal:trans_iso_T}. } 

\hl{It has to be noted that in this thermo-poro-elastic framework, we neglect possible reversible sorption-desorption phenomena, which have been widely discussed on certain geomaterials such as clay or coal (e.g. \mbox{\citealp{Martin196228,Liu201845}}).} 
%Similar to the work of \cite{Ma199215}, we can assume that the solid phase includes mineral constituents and strongly adsorbed water, while the fluid phase comprises only (free) water. 
\hl{Here we do not consider a mass exchange between water adsorbed within the solid phase and bulk water in the pore space.}
\hl{The vector of thermal expansion} $\alpha_{d,i}$ consists of the two linear drained thermal expansion coefficients perpendicular ($\alpha_{d,z}$) and parallel ($\alpha_{d,h}$) to the bedding orientation: 
\begin{equation}
\label{eq:thermal:advector}
\alpha_{d,i} = {\left[ {{\alpha_{d,h}},{\alpha_{d,h}},{\alpha_{d,z}},0,0,0} \right]^\top}
\end{equation}

The volumetric drained thermal expansion coefficient $\alpha_{d}$ is calculated with $\alpha_{d}=2\alpha_{d,h}+\alpha_{d,z}$.
\hl{The volumetric strain response of the porous medium due to pore pressure change under constant total stress and temperature is characterized by the Biot modulus $H$:}

\begin{equation}
\frac{1}{H} = -{\left( {\frac{{\rm{d}}\varepsilon_{v}}{{{\rm{d}} {p_f}}}} \right)_{{\rm{d}\sigma},\mathrm{d}T}} = \frac{1}{{{K_d}}} - \frac{1}{{{K_s}}}
\label{eq:thermal:2}
\end{equation}
where $\varepsilon_{v}$ is the volumetric strain, $K_d$ the isotropic drained bulk modulus and $K_s$ unjacketed \hz{bulk} modulus \hz{\mbox{\citep{Brown1975}}. The modulus $K_s$ relates volumetric strains to stress changes, when equal increments of isotropic stress and pore pressure are applied to a specimen.} The Biot modulus can also be expressed in terms of components $H_i$ in different anisotropy directions $i=z,h$, by using:

\begin{equation}
\frac{1}{H_i} = -{\left( {\frac{{\rm{d}}\varepsilon_{i}}{{{\rm{d}} {p_f}}}} \right)_{{\rm{d}\sigma},\mathrm{d}T}}   
\label{eq:thermal:3}
\end{equation}

\hl{A relationship between the volumetric modulus $H$ and the linear moduli $H_i$ is found by $H=2H_h+H_z$.}
\hl{Undrained conditions are defined for a constant pore} fluid mass in the \hl{porous medium under loadings. Under this condition}, one obtains the following \hl{changes} in strains \hl{under} constant stresses: 
\begin{equation}
\label{eq:thermal:epsiu}
{\rm{d}} \varepsilon_{i}= -\alpha_{u,i}{\rm{d}}T
\end{equation}
where $\alpha_{u,i}$ denotes the two linear undrained thermal expansion coefficients perpendicular (${\alpha_{u,z}}$) and parallel ($\alpha_{u,h}$) to the bedding orientation:
\begin{equation}
\label{eq:thermal:auvector}
{             \alpha_{u,i} = {\left[ {{\alpha_{u,h}},{\alpha_{u,h}},{\alpha_{u,z}},0,0,0} \right]^\top}            } 
\end{equation}
The volumetric undrained thermal expansion coefficient can be calculated according to $\alpha_{u}=2\alpha_{u,h}+\alpha_{u,z}$.
The pore pressure changes in undrained conditions under constant total stress are described by:
\begin{equation}
\label{eq:thermal:pu}
{     {\rm{d}}{p_f} = \Lambda{\rm{d}}T			}
\end{equation}
where $\Lambda$ represents the thermal pressurization coefficient \citep{Ghabezloo200954}. \hl{According to \mbox{\cite{Ghabezloo200954}} and by using Eq. (\ref{eq:thermal:2}), $\Lambda$} can be expressed as:
\begin{equation}
\label{eq:thermal:lambda} 
\Lambda  = \frac{{\left( {{\alpha _f} - {\alpha _\phi }} \right)}}{{\frac{1}{\phi H} +  \left( {\frac{1}{{{K_f}}} - \frac{1}{{{K_\phi }}}} \right)}}
\end{equation}
where $\alpha _f$ and $\alpha _\phi$ are the bulk thermal expansion coefficients of the pore fluid of the pore space, respectively. $K_\phi$ denotes the \hl{unjacketed pore modulus \mbox{\citep{Brown1975}},} $\phi$ the porosity, $H$ the Biot's moduls and $K_f$ the fluid bulk modulus. 

Eq. (\ref{eq:thermal:lambda}) \hl{shows} that the thermal pressurization is mainly caused by the difference between the bulk thermal expansion coefficients of the fluid $\alpha _f$ and that of the pore space $\alpha _\phi$. In geomechanical applications, when the material is micro-homogeneous, we can assume that $\alpha _\phi$ is equal to the drained volumetric thermal expansion coefficient $\alpha _d$. In the same manner, one can take the bulk modulus of the pore space $K_\phi$ equal to the unjacketed bulk modulus $K_s$.  

Comparing Eq. (\ref{eq:thermal:epsi}) with (\ref{eq:thermal:epsiu}) and (\ref{eq:thermal:pu}), one obtains the following relationship:
\begin{equation}
\label{eq:thermal:relu}
\alpha_{u,i}=\frac{\Lambda}{H_i} +\alpha_{d,i}    
\end{equation}
This can be simplified in terms of \hl{the bulk volume change} \citep{Ghabezloo200954} as follows:
\begin{equation}
\label{eq:thermal:reluiso}
\alpha_{u}=\frac{\Lambda}{H} +\alpha_{d}    
\end{equation}
We can evaluate $\alpha_{u,z}$, $\alpha_{u,h}$ and ${\Lambda}$ in undrained heating tests, whereas $\alpha_{d,z}$ and $\alpha_{d,h}$ are determined in drained heating. The moduli $H_{z}$ and $H_{h}$ can then be calculated using Eq. (\ref{eq:thermal:relu}). Alternatively, if $H_{i}$ are \hl{also} measured, the parameter set ($\alpha_{u,i}$, ${\Lambda}$, $H_{i}$ and $\alpha_{d,i}$) \hl{is overdetermined, and the values} can be checked for \hl{their consistency} (i.e. for each unknown we can compare its measured value with the one back-calculated from the other unknowns, see also \citealp{Menke1989,Hart199574,Braun2019}).

\section{Materials and Methods}
\label{sec:thermal:3}

\subsection{The Callovo-Oxfordian Claystone}
\label{sec:thermal:3.1}
Samples of the COx claystone have been retrieved at Bure, in the east of France, from an underground research laboratory (URL). This URL at about 490 m depth was established by the French national agency for radioactive waste management, Andra, to study the behaviour of the claystone during tunnel excavation and during the long term disposal of radioactive waste \hl{(\mbox{\citealp{Andra2005}}; \mbox{\citealp{Armand201741}}; \mbox{\citealp{Conil201861}}). 
At the URL level, the claystone has a clay content of around 42 $\%$, with an average porosity of 17.5 $\%$ and an average water-content of 7.9 $\%$ \mbox{\citep{Conil201861}}.}
\hl{The solid matrix comprises $10–24 \%$ interstratified illite/smectite layers, $17–21 \%$ illite, $3–5 \%$ kaolinite and $2–3 \%$ chlorite \mbox{\citep{Gaucher200455}}.}
\cite{Wileveau200786} determined a vertical and a minor horizontal total stress of around 12 MPa, a major horizontal stress of around 16 MPa and a pore pressure of about 4.9 MPa. 

\begin{figure}[htbp]
% Use the relevant command to insert your figure file.
% For example, with the graphicx package use
\centering
  \includegraphics[width=0.5\columnwidth]{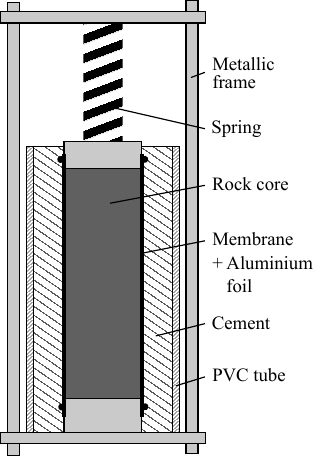}%{cell.pdf}
% figure caption is below the figure
\caption{\hl{Schematic cross-section of a T1 cell \mbox{\citep{Conil201861}}, showing a cylindrical rock core within different layers of protection}.}
\label{fig:thermal:T1cell}       % Give a unique label
\end{figure}

In this laboratory study, we investigated specimens trimmed from COx cores that were extracted at the URL from horizontal boreholes.
To preserve the material in-situ state as best as possible, i.e. to avoid desaturation \citep{Monfared201173,Ewy201538} and mechanical damage, cores were shipped and stored in T1 cells \hl{(Fig. \ref{fig:thermal:T1cell}),} developed by Andra \citep{Conil201861}. T1 cells protect 80 mm diameter COx cores that are wrapped in aluminium foil and covered by latex membranes to prevent drying. A PVC tube is placed around the core and cement is poured between the membrane and the tube. Once hardened, the cement ensures radial confinement and shock protection. A mechanical spring is installed, which applies confining stress in axial direction.

After having removed the cores EST 53650 and EST 57185 from the T1 cells, the cores were sealed by a layer of paraffin wax. This layer provided drying protection and remained during the subsequent trimming process. Cylinders of 38 mm diameter were then drilled perpendicular to the bedding plane, using a air-cooled diamond coring bit. The cylinders were also protected by a paraffin layer, before being cut into 10 mm thick disks with a diamond string saw. The disks were enveloped by an aluminium-foil layer and covered by a mixture of 70 \% paraffin wax and 30 \% vaseline oil. 

A petrophysical characterization, performed on cuttings of the cores directly after trimming and several months after storage, under the before-mentioned protection, is presented in Tab. \ref{tab:thermal:1}. The \hl{specimen} volume was determined by hydrostatic weighting in petroleum, while the dry density was obtained after oven drying at 105 $^{\circ}$C for two days. A value for the solid density $\rho_s=$ 2.\hl{69} g/cm$^3$ was provided by Andra \hl{\mbox{\citep{Conil201861}}}. The measurements of suction $s$ were carried out using a chilled mirror tensiometer (WP4, Decagon brand). We observed a relatively high degree of saturation, \hl{larger than 92.5 \%,} confirming a good sample preservation.

As indicated in Tab. \ref{tab:thermal:test_overview}, four \hl{specimens}, ISO1 to ISO4, were tested in this study. Sample ISO1 was trimmed from core EST53650 and samples ISO2 to ISO4 from core EST57185.
%
% For tables use
\begin{table}
% table caption is above the table
\caption{Mean and standard deviation (in parentheses) of petrophysical measurements done on cuttings of two COx cores.}
\label{tab:thermal:1}       % Give a unique label
\begin{tabular}{ lcccccc }
	\hline\noalign{\smallskip}
  	\multirow{2}{*}{Core}	& {$\rho$} & {$\rho_d$ } & {$\phi$} & {$w$} & {$S_r$}	& {$s$}\\
  							& $[\text{g/cm}^3]$ & $[\text{g/cm}^3]$  & $[\%]$ & $[\%]$ & $[\%]$	& $[\text{MPa}]$\\
	\noalign{\smallskip}\hline\noalign{\smallskip}
 	{EST} 		& 2.37 		& 2.22 		& 17.9 		& 7.5 		& 92.5 		& 24.2\\
 	{53650}		& (0.00) 	& (0.01) 	& (0.2)  	& (0.1) 	& (0.8)		& (2.1)\\
 	\noalign{\smallskip}\hline\noalign{\smallskip} 
 	{EST} 		& 2.38 		& 2.21 		& 18.2 		& 7.9 		& 95.3 		& 17.4 \\
 	{57185}		& (0.00) 	& (0.00) 	& (0.2)  	& (0.1) 	& (0.7)		& (0.1)	\\
\noalign{\smallskip}\hline
\end{tabular}
\end{table}

\subsection{Isotropic cell}
\label{sec:thermal:3.2}

\begin{figure}[htbp]
% Use the relevant command to insert your figure file.
% For example, with the graphicx package use
  \includegraphics[width=\columnwidth]{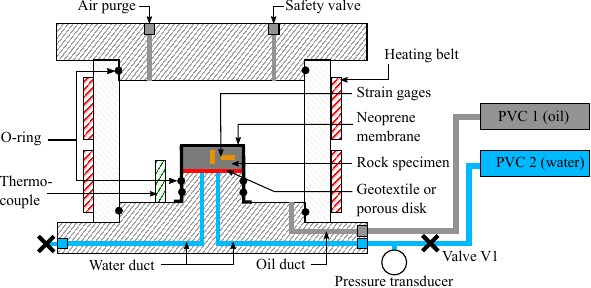}%{cell.pdf}
% figure caption is below the figure
\caption{Temperature controlled isotropic compression cell, accommodating disk shaped rock specimens equipped with strain gages. \hl{A heating belt connected with a thermocouple regulates the cell temperature. An external oil pump controls the isotropic confining pressure, while a water pump allows saturation and back pressure application. Valve V1 permits one to close the drainage system, creating undrained conditions, while the pore pressure transducer records the fluid pressure.}}
\label{fig:thermal:cell}       % Give a unique label
\end{figure}

A high-pressure isotropic thermal compression cell, presented in Fig. \ref{fig:thermal:cell} \citep{Tang200845,Mohajerani201211,Belmokhtar201722,Belmokhtar201787}, was used to test specimens with 38 mm diameter and variable height, sealed within a neoprene membrane, in the centre of the cell. A \hl{silicone rubber electric} heating belt \hl{is wrapped around the cell and} coupled with a temperature sensor next to the specimen, \hl{giving control over the specimen temperature}. The cell is filled with silicone oil, which can be set under a pressure of up to 40 MPa by a pressure volume controller (PVC 1, GDS brand). The neoprene membrane isolates the specimen from the silicone oil, so that the pressure of the pore fluid can be controlled independently. The pore pressure is applied from the bottom of the specimen through a porous disk connected to the pressure volume controller PVC 2. Local strains wereto the data of monitored using an axial and a radial strain gage (Kyowa brand) glued on the lateral surface of the specimen.

\subsection{Correction of strain measurements}
\label{sec:thermal:correction}
\hl{Measurements of strain gages can be altered by both temperature changes of the gage itself inside the cell and changes of the room temperature, affecting external cables. To account for these effects, a reference gage glued on a piece of steel 316L was placed next to the specimen. Knowing the thermal expansion coefficient of steel 316L, the thermally induced error strains, called apparent strains $\varepsilon_T$, can be measured simultaneously to the specimen strains and allow for real-time corrections. We found apparent strains in the same order of magnitude as the specimen strains, which emphasizes the necessity for correction.} 

\begin{figure}[htbp]
\includegraphics[width=\columnwidth]{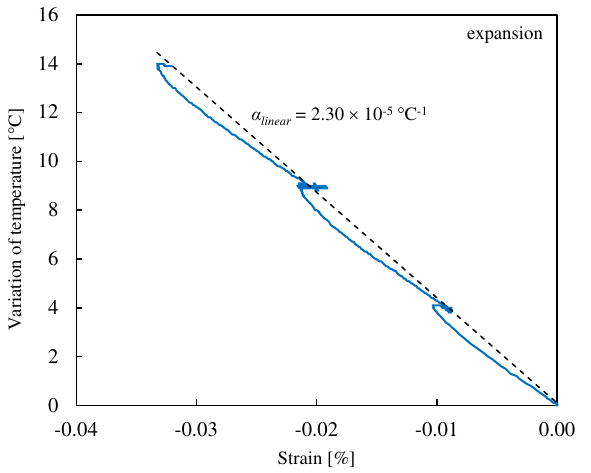}
\caption{\hl{Validation of the thermal strain measurement through a heating test on} an aluminium 2011 specimen, heated from 26 ${^\circ \mathrm{C}}$. \hl{The obtained linear thermal expansion coefficient $\alpha_{\rm{linear}}$ corresponds with the value given by the manufacturer.}}
\label{fig:thermal:alu}       % Give a unique label
\end{figure}

To verify the thermal strain measurements, a specimen of aluminium 2011 was equipped with a strain gage and tested in the cell. The reference gage was placed next to the specimen. The cell was heated in steps from 26 - 30 - 35 - 40 $^{\circ}$C, while the strains of the aluminium \hl{specimen} were recorded and corrected by $\varepsilon_T$. Fig. \ref{fig:thermal:alu} shows the linear thermal expansion of the aluminium, with $\alpha_{\rm{linear}}=2.30 \times 10^{-5}  {^\circ \mathrm{C}}^{-1}$. This value corresponds to the value given from the manufacturer, \hl{validating a precise temperature correction}.  

Among the four tests carried out, a reference gage was installed in two tests (ISO3 and ISO4, see Tab. \ref{tab:thermal:test_overview}). In the two other tests (ISO1 and ISO2), the measured specimen strains were corrected a posteriori with a calculated $\varepsilon_T$, determined in function of cell temperature and room temperature.

\subsection{Calibration of the effect of the drainage system}
\label{sec:thermal:calibration}

According to \cite{Wissa196910,Bishop197637} and \cite{Ghabezloo200901,Ghabezloo201060}, undrained conditions \hl{imposed by a testing device} might not always \hz{meet} the requirement of a constant fluid mass inside the specimen. If there is some dead volume in the drainage system, such as ducts and porous disks, this volume changes due to the compressibility and thermal expansion of the drainage elements and of the fluid it contains. This allows fluid to either drain or be injected into the specimen, changing the fluid mass. The methods proposed to account for these effects require a precise calibration of the testing device. Using  Eq. (\ref{eq:thermal:corrLambda}), presented by \cite{Braun2019} and adapted from \cite{Ghabezloo201060}, the corrected thermal pressurization coefficient $\Lambda^{\rm{cor}}$ is obtained from the measured one $\Lambda^{\rm{mes}}$:
\begin{equation}
\label{eq:thermal:corrLambda}
\Lambda ^{\rm{cor}} = \frac{{{\Lambda ^{\rm{mes}}}}}{{1 - \frac{{{V_L}\left( {{c_f} + {c_L}} \right)}}{{\phi V\left( {{\alpha _f} - {\alpha _d }} \right)}}\left( {{\Lambda ^{\rm{mes}}} - {\Lambda _L}} \right)}}
\end{equation}
\hl{where $V_L$ is the volume, $c_L$ the compressibility and $\Lambda_L$ the thermal pressurization coefficient of the drainage system. $\phi$ is the porosity and $V$ the total volume of the specimen. $c_f = 1/K_f$ denotes the pore fluid compressibility.} For sake of simplicity, we consider an ideal porous material with $\alpha_{\phi} = \alpha_{d}$. The \hl{parameter $\Lambda_L$ of the testing device} is determined by running a calibration test on a dummy metal specimen with zero porosity. Also, $\phi$ and $V$ of \hl{each} specimen have to be \hl{measured before a} test.

Similarly, we calculate a corrected undrained thermal expansion coefficient $\alpha _u^{\rm{cor}}$ with the measured one $\alpha _u^{\rm{mes}}$, which requires knowledge of the Biot modulus $H$ \citep{Ghabezloo201060}:
\begin{equation}
\label{eq:thermal:corralpha}
\alpha _u^{\rm{cor}} = {\alpha _d} + \frac{{\left( {\alpha _u^{\rm{mes}} - {\alpha _d}} \right)}}{{1 - \frac{{{V_L}\left( {{c_f} + {c_L}} \right)}}{{\phi V\left( {{\alpha _f} - {\alpha _d }} \right)}}\left[ {\left( {\alpha _u^{\rm{mes}} - {\alpha _d}} \right)H - {\Lambda _L}} \right]}}
\end{equation}

In this study, calibration was \hl{done} by adopting $V_L = 2000$ mm$^3$, $c_L = 0.30$ GPa$^{-1}$ and a temperature dependent $\Lambda_L  =$ (0.0056 $T$ + 0.208) $\rm{MPa} ^{\circ} \rm{C} ^{-1} $\hl{, with $T$ in $^{\circ} \rm{C}$ (see \mbox{\citealp{Braun2019}}, who used a similar device). For the fluid compressibility $c_f = 1/K_f$ we used properties of bulk water, taking into account its temperature and pressure dependency \mbox{\citep{IAPWS-IF972008}}.}  

\subsection{Testing Procedure}
\label{sec:thermal:3.3}
\begin{figure}[htbp]
\includegraphics[width=\columnwidth]{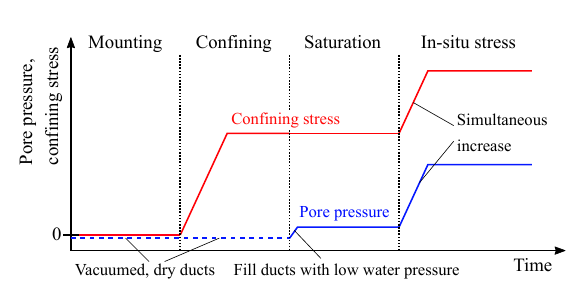}
\caption{\hl{Mounting and saturation procedure for COx specimens. Specimens are first mounted with dry water-ducts. Vacuum is applied to the ducts in order to evacuate air. The isotropic confining stress is then increased and kept constant. When monitored strains are stable, the water saturation is started with a low back pressure to limit poroelastic strains. After the hydration swelling stabilized, both pore pressure and confining stress are increased simultaneously until the desired stress level close to the in-situ one is reached. Again, strains are monitored and when the deformations are in equilibrium, further testing can be started.}}
\label{fig:thermal:saturation_procedure}       % Give a unique label
\end{figure}

Before installing a specimen in the testing device, the drainage ducts were \hl{emptied and dried by flushing with compressed air. Any contact of the claystone with water, before applying confining stress, could lead to swelling and risk specimen damage (\mbox{\citealp{Mohajerani201160}}; \mbox{\citealp{Menaceur201529}}; \mbox{\citealp{Ewy201538}}). Following the procedure illustrated in Fig. \ref{fig:thermal:saturation_procedure}, the} COx specimens were then mounted in the isotropic cell in their initial state at a constant ambient temperature. The specimens were consolidated \hl{without back pressure application} under isotropic stress loading of 0.1 MPa/min. 

All specimens were brought to a confining pressure close to the in-situ effective stress (sample ISO1 to 10 MPa total stress and samples ISO2 - ISO4 to 8 MPa total stress, see Tab. \ref{tab:thermal:test_overview}) while monitoring deformations. Once the consolidation deformations stabilized, vacuum was applied to the drainage ducts to evacuate the remaining air. The dry ducts were then filled with synthetic pore water with a composition close to the in-situ one, provided by Andra, to saturate the specimens.

\begin{table*}
\caption{Characteristics of the four tested specimens ISO1 - ISO4, with the applied isotropic confining stress $\sigma$ and measured volume change $\varepsilon_{\mathrm{hyd}}$ during saturation. The utilized testing procedure with imposed confining stress $\sigma$ and pore pressure $p_f$, and the average measured thermo-elastic parameters are also shown (note that thermo-plastic characteristics and temperature dependencies are not detailed here).}
\label{tab:thermal:test_overview}     % Give a unique label
\begin{tabular}{@{\extracolsep{0pt}}llccccccccccc@{}}
\hline\noalign{\smallskip} 
       &          & \multicolumn{2}{l}{Saturation}    & \multicolumn{3}{l}{Test conditions}  & \multicolumn{6}{l}{Average measured parameters}  \\
\cline{3-4}\cline{5-7}\cline{8-13}\noalign{\smallskip}
 & Core & $\sigma$ & $\varepsilon_{\mathrm{hyd}}$ & Type       & $\sigma$ & $p_f$ & $\alpha_{d,z}\!^{(1)}$   & $\alpha_{d,h}$ & $\alpha_{u,z}\!^{(2)}$ & $\alpha_{u,h}\!^{(2)}$ & $\Lambda\!^{(2)}$ & $H\!^{(2)}$  \\
\# & EST & {[}MPa{]}              &  {[}\%{]}   &          & {[}MPa{]}              & {[}MPa{]}           & \multicolumn{4}{c}{ $[10^{-5}{^\circ \mathrm{C}}^{-1} ]$ }         & {[}MPa/$^{\circ}$C{]}    & {[}GPa{]}  \\
\noalign{\smallskip} \hline\noalign{\smallskip} 
ISO1   & 57185    & 10               & 0.71     & three-stage                  & 14               & 4             & -                           & 0.53      & 4.82        & 2.30        & 0.19     & 2.32 \\
ISO2   & 53650    & 8                & 0.86     & two-stage                      & 12               & 4             & -                           & 0.60      & -           & 2.03        & -        & -    \\
ISO3   & 53650    & 8                & 0.80     & two-stage                     & 12               & 4             & 0.18                        & 0.30      & 4.63        & 1.86        & -        & -    \\
ISO4   & 53650    & 8                & 0.72     & three-stage                      & 14               & 4             & 0.22                        & 0.51      & 4.93        & 1.99        & 0.20     & 2.49 \\
\noalign{\smallskip} \hline\noalign{\smallskip} 
\multicolumn{13}{l}{$^{(1)}$only thermo-elastic values shown here}                                                                                                                                                      \\
\multicolumn{13}{l}{$^{(2)}$temperature dependent}                                                                                                                                                                     
\end{tabular}
\end{table*}

The fluid pressure was chosen small enough in order to limit the change of effective stress \hl{and related poroelastic strains}. Some swelling strains were observed that stabilized after about 5 days, with 0.71 \% expansion for ISO1, 0.86 \% expansion for ISO2, 0.80 \% expansion for ISO3 and 0.72 \% expansion for ISO4 (see Tab. \ref{tab:thermal:test_overview}). These values are comparable to the data of \cite{Belmokhtar201787}, who found 0.97 and 0.85 \% for \hl{specimens} saturated at 8 and 12 MPa effective stress, respectively. 

After hydration, pore pressure and confining pressure were simultaneously increased under constant effective stress, until reaching pore pressure close to the in-situ one (4.0 MPa). \hl{This back pressure seemed sufficient for complete saturation, according to \mbox{\cite{Rad1984}}. They showed, that with the help of vacuum, a specimen at 90 \% saturation can be fully re-saturated under around 150 kPa pore pressure. The key advantage is hereby the dissolution of air into water under pressure, according to Henry's law. Also \mbox{\cite{Favero2018}} demonstrated that their specimens of Opalinus clay were completely saturated under 2 MPa pore pressure. Our samples} ISO1 and ISO4 were brought to 14 MPa total stress, and samples ISO2 - ISO3 to 12 MPa total stress. Note that an \hl{additional} increment of 2 MPa confining pressure was finally applied to ISO4, resulting in higher effective stress than during saturation.

\begin{figure*}[htbp]
% Use the relevant command to insert your figure file.
% For example, with the graphicx package use
  \includegraphics[width=\textwidth]{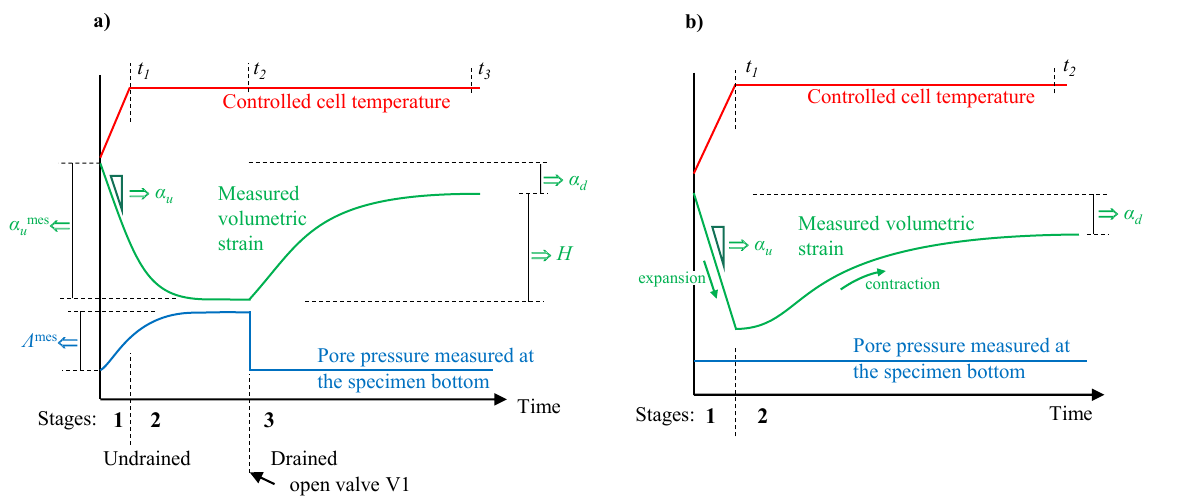}
% figure caption is below the figure [width=0.75\textwidth]
\caption{Schematic procedure of a) three-stage thermal test and b) two-stage thermal test. \hl{In both tests, pore pressures are generated within the specimen due to heating. In a), the drainage system is closed first. The generated pore pore can equilibrate with the drainage system and be measured by the external sensor. Afterwards, the overpressure is released. In b), the drainage system is kept open. Nonetheless, thermally induced pore pressures are created within the specimen during loading, but are able to drain directly after loading is stopped. One can measure strains, but no external pore pressure measurement can be done. }}
\label{fig:thermal:procedure}       % Give a unique label
\end{figure*}

Two different step testing procedures, presented by \cite{Braun2019}, were utilized to determine the THM parameters of the material. 
Samples ISO1 and ISO4 were tested following a three-stage protocol (Fig. \ref{fig:thermal:procedure}a), that consists in rapidly increasing or decreasing the cell temperature \hz{in stage 1 until the time $t_1$} (rate of 10 $^{\circ}$C/h), while the drainage system is kept closed. In this phase, one can measure \hz{$\alpha_{u}$ (with its anisotropic components $\alpha_{u,i}$)} \hl{based on the temperature and strain measurements, using the initial tangent of $\alpha_{u}=-{\mathrm{d}\varepsilon_v}/{\mathrm{d}T}$. Initially, no pore pressure dissipation occurs, so that there is no} need of correcting any effect of the drainage system \hl{\mbox{\citep{Braun2019}}.} In \hz{stage 2 until $t_2$}, the temperature is kept constant until deformations and pore pressures stabilized. The corresponding pseudo-undrained conditions require the correction of the measured secant parameter \hz{$\alpha_{u}^{mes}=-{\mathrm{d}\varepsilon_v}/{\mathrm{d}T}$ and $\Lambda^{mes}=\mathrm{d}p_f/\mathrm{d}T$}, according to Eqs. (\ref{eq:thermal:corrLambda}) and (\ref{eq:thermal:corralpha}). In \hz{stage 3}, the drainage system is opened, imposing the initial pore pressure. The pore pressure gradually comes back to its initial distribution through a transient phase governed by pore pressure diffusion, and a drained state is finally reached. \hz{At $t_3$}, after strain stabilization, \hz{the secant parameter $\alpha_{d}=-{\mathrm{d}\varepsilon_v}/{\mathrm{d}T}$} can be determined. The change in pore pressure with respect to the change of strains during the third phase provides the \hz{secant modulus $H=-\mathrm{d}p_f/\mathrm{d}\varepsilon_v$.}
 
Samples ISO2 and ISO3 were tested under a two-stage procedure (Fig. \ref{fig:thermal:procedure}b). The \hl{specimens} are first rapidly heated or cooled \hz{in stage 1 until $t_1$,} under 10 $^{\circ}$C/h, and then kept at constant temperature \hz{in stage 2}. Even though the drainage system is always kept open, the temperature change is much faster than pore pressure diffusion, and one can measure \hz{the undrained tangent parameter $\alpha_{u}=-{\mathrm{d}\varepsilon_v}/{\mathrm{d}T}$}. After reaching stable strains in a completely drained state \hz{at $t_2$}, \hz{the secant parameter $\alpha_{d}=-{\mathrm{d}\varepsilon_v}/{\mathrm{d}T}$} can be measured. 
\hz{Note that in both procedures, for each measurement of the volumetric properties $\alpha_{u},\alpha_{u}^{mes}, \Lambda^{mes}, \alpha_{d}$ and $H$, we obtain at the same time the corresponding linear components with respect to the two directions parallel and perpendicular to bedding, e.g. $\alpha_{u}$ with $\alpha_{u,h}$ and $\alpha_{u,z}$.}

The isotropic total stress was always kept constant during the tests, and the pore pressure at the bottom of the specimen was \hz{kept} constant by the PVC. \hz{Even though the pore pressure at the bottom is constant, the pore pressure distribution within the specimen changes during thermal loading due to thermal pressurization. However, at the end of the step tests after pore pressure dissipation, the specimen is in hydraulic equilibrium, resulting in a constant pore pressure distribution equal to the initial pore pressure}. The stress conditions before and after the step test, when drained strains are measured, are identical. Drained thermal strain changes were therefore always evaluated under constant stresses.
\hl{In the intermediate undrained phase however, the pore pressure varies according to ${\rm{d}}p_f=\Lambda{\rm{d}}T$ (Eq. (\ref{eq:thermal:pu})). Consequently, this changes the effective stress depending on the increment of temperature applied in one step and the thermal pressurization coefficient. If the material properties are stress dependent, this change of effective stress has to be considered in the analysis of undrained parameters. Also, a temperature dependency of the observed properties has to be taken into account, if the temperature increments are large.} 

\begin{figure}[htbp]
% Use the relevant command to insert your figure file.
% For example, with the graphicx package use
  \includegraphics[width=\columnwidth]{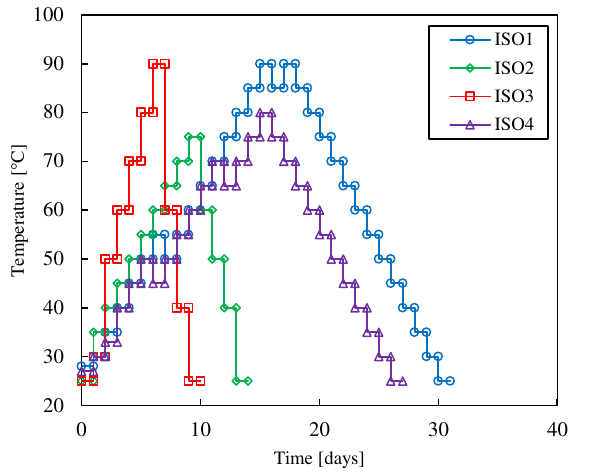}
% figure caption is below the figure [width=0.75\textwidth]
\caption{Applied heating and cooling paths on the four tested \hl{specimens. In} each step the temperature was changed rapidly with increments from 5 $^{\circ}$C (ISO1, ISO4) up to 15 $^{\circ}$C (ISO2, ISO3), followed by a constant temperature stage during one day.}
\label{fig:thermal:heatingpath}       % Give a unique label
\end{figure}

The temperature, at which the measured parameters are presented in the following, was chosen as the mean between initial and final temperatures during each step test.  
\hl{In a series of subsequent temperature step tests, the specimens were heated until 90 $^{\circ}$C (ISO1, ISO3), 80 $^{\circ}$C (ISO4) and 75 $^{\circ}$C (ISO2), and then cooled back, as presented in Fig. \ref{fig:thermal:heatingpath}.}
The utilized testing procedures, as well as stress and pore pressure conditions and measured material parameters are summarized in Tab. \ref{tab:thermal:test_overview} (note that the axial strain gage on sample ISO2 failed after saturation, so that only radial strains could be recorded during this test).

\section{Experimental results}
\label{sec:thermal:4}
\begin{figure}[htbp]
  \includegraphics[width=\columnwidth]{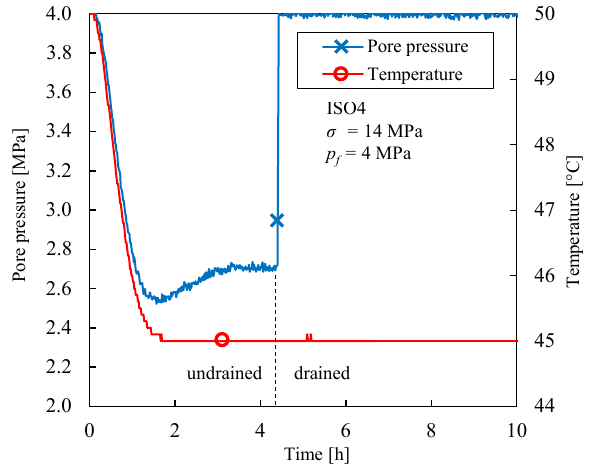}
\caption{Typical three stage thermal test on ISO4. \hl{First we observe a decrease} in pore pressure resulting from an undrained decrease in temperature form 50 to 45 $^{\circ}$C. \hl{The drainage system is then opened, re-imposing the initial pore pressure at the specimen bottom.}}
\label{fig:thermal:expp}      
\end{figure}

\begin{figure}[htbp]
  \includegraphics[width=\columnwidth]{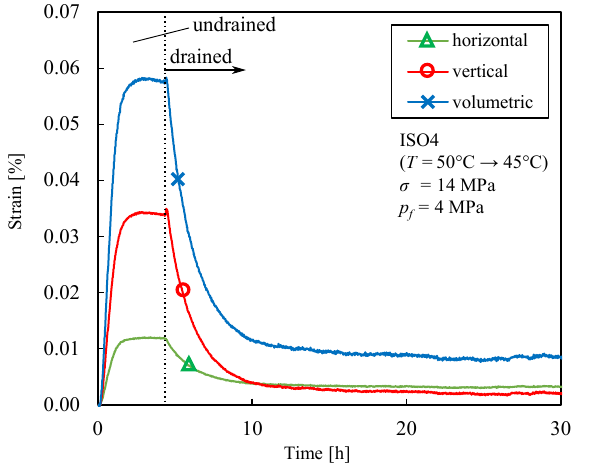}
\caption{Measured strains parallel and perpendicular to the bedding plane during test ISO4, with an initial undrained thermal expansion stabilizing after a 4 hours. \hl{Afterwards we re-impose the initial pore pressure. As the pore pressure within the specimen at this point is lower than the one at the bottom, pore pressure dissipates into the specimen and induces a transient expansion. The strains stabilize in a completely drained state.}}
\label{fig:thermal:exstrain}      
\end{figure}

The data of a typical three-stage thermal test, run on specimen ISO4 between 50 to 45 $^{\circ}$C, \hl{are} presented in Fig. \ref{fig:thermal:expp} and \ref{fig:thermal:exstrain} in terms of pore pressure and strain changes, respectively. \hz{Here, the specimen was cooled and the temperature decrease resulted in a pore pressure decrease (thermal de-pressurization during cooling, equivalent to thermal pressurization during heating).} The target temperature was reached after about 1.5 h and then kept constant. Given that the drainage system was closed, a decrease in pore pressure due to thermal \hl{de-pressurization} is observed, which stabilizes after 4 h. The peak at 1.5 h is due to the thermal pressurization coefficient of the drainage system $\Lambda_L$, which was found to be higher than that of the specimen. As also observed by \cite{Mohajerani201211}, due to the rapid temperature change, there is initially no exchange between the fluid of the drainage system and the pore fluid of the specimen. One only detects the response due to the \hl{de-pressurization} of the drainage system itself. After about 4 h, the pressure equilibrates with the specimen pore pressure. The measured thermal pressurization and expansion coefficient can then be corrected using Eq. (\ref{eq:thermal:corrLambda}) and (\ref{eq:thermal:corralpha}), respectively. To avoid uncertainties \hl{related to this calculation}, we \hl{preferred} to determine $\alpha_{u,i}$ from the initial slope of -d$\varepsilon_i$/d$T$ instead. This \hl{measurement provides directly} the undrained thermal expansion coefficients in both anisotropy directions\hl{, with no need of correction \mbox{\citep{Braun2019}}.}
Once the equilibrium in undrained state was reached, we opened the drainage valve, so that the pore pressure in the drainage system and at the specimen bottom increased \hl{instantaneously} back to its initial value\hl{. The difference between specimen pore pressure and constant boundary fluid pressure is now gradually equilibrating through fluid dissipation. The pore fluid dissipates into the specimen, the pore pressure increases and expansive strains are recorded, similar to a transient pore pressure test (e.g. \mbox{\citealp{Ghabezloo200976}}).} Such behaviour was also observed in thermal test on Boom clay by \cite{Delage200034}. The pore pressure dissipation observed in the present study caused transient deformations, that stabilized after about 20 h. This deformations can be seen in Fig. \ref{fig:thermal:exstrain}, presenting the changes in horizontal, vertical and volumetric strains with respect to time. Given that complete drainage was achieved, the drained thermal expansion coefficients were measured. The Biot modulus $H=-\Delta p_f / \Delta \varepsilon_v$ could be computed from the volumetric deformations, under the pore pressure change of 1.3 MPa between the undrained and the drained phase. 
Once this temperature step completed, the drainage system was closed again and a subsequent similar three stage test was carried out.

In the two-stage tests on samples ISO2 and ISO3, we went directly from rapid thermal loading to a transient drained phase. The measured parameters and the characteristics of the transient deformations obtained in these phases are comparable to that from three-stage tests.  

\subsection{Drained thermal deformations}
\label{sec:thermal:4.2}

\begin{figure}[htbp]
  \includegraphics[width=\columnwidth]{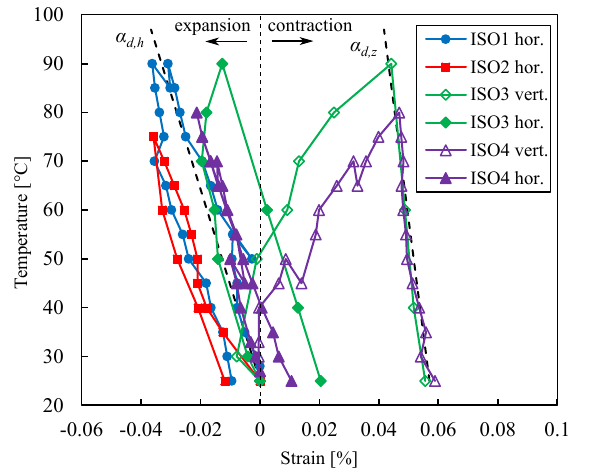}
% figure caption is below the figure [width=0.75\textwidth]
\caption{Drained strain changes (vertical = perpendicular and horizontal = parallel to bedding) with respect to temperature changes, under constant isotropic effective stress (10 MPa for ISO1 and ISO4 and 8 MPa for ISO2 and ISO3).}
\label{fig:thermal:drdefh}       % Give a unique label
\end{figure}

At the end of each \hl{temperature} step, a drained state was attained when deformations stabilized. These strain changes with respect to temperature are presented in Fig. \ref{fig:thermal:drdefh}. The drained deformations in horizontal direction parallel to the bedding plane were recorded for all four samples. Specimens ISO1 and ISO2 showed some \hl{measurement} artefacts in \hz{the} vertical direction perpendicular to the bedding plane and \hl{are} not presented here.

Strains parallel to bedding were observed to be predominantly reversible and reasonably linear with temperature. In that direction, specimens expanded during heating and contracted during cooling, illustrating a thermo-elastic response with an average linear elastic thermal expansion coefficient $\alpha_{d,h}=0.51 \times 10^{-5}  {^\circ \mathrm{C}}^{-1}$.

Perpendicular to bedding, the material first expanded with temperature up to 40 ${^\circ \mathrm{C}}$, followed by contraction at higher temperatures. During cooling, one can observe a linear contraction, providing an average linear thermal expansion coefficient $\alpha_{d,z}=0.21 \times 10^{-5}  {^\circ \mathrm{C}}^{-1}$ from tests ISO3 and ISO4. Note that this thermal expansion coefficient is about 2.4 times smaller than that measured parallel to bedding. This gives a bulk thermal expansion coefficient $\alpha_{d}=1.23 \times 10^{-5}  {^\circ \mathrm{C}}^{-1}$.  
  
To describe the observed contraction during heating, we use an elasto-plastic drained thermal expansion coefficient $\alpha_{d,z}^* = \alpha_{d,z} + \alpha_{d,z}^{\rm{irr}}$, which includes reversible ($\alpha_{d,z}$) and irreversible ($\alpha_{d,z}^{\rm{irr}}$) thermal strains \citep{Sulem200752}. We can adopt a linear relationship from the measurements on samples ISO3 and ISO4, \hl{found by a least square error fitting on the data in Fig. \ref{fig:thermal:drdefh}:}
\begin{equation}
\label{eq:thermal:}
\alpha_{d,z}^* =\left(-0.035 T +1.238 \right) \times 10^{-5}  {^\circ \mathrm{C}}^{-1}    \text{,}
\end{equation}
with $T$ in ${^\circ \mathrm{C}}$ \hl{from 25 to 90 ${^\circ \mathrm{C}}$}, providing the value of the irreversible thermal expansion coefficient:
\begin{equation}
\label{eq:thermal:}
\alpha_{d,z}^{\rm{irr}} =\left(-0.035 T +1.071 \right) \times 10^{-5}  {^\circ \mathrm{C}}^{-1} 
\end{equation}

The contraction behaviour under heating was previously observed by \cite{Monfared201173} on the Opalinus claystone and by \cite{Belmokhtar201722} on the COx claystone.
\cite{Belmokhtar201722} determined $\alpha_{d}=4.82 \times 10^{-5} {^\circ \mathrm{C}}^{-1}$, by means of LVDTs, a value much larger than that observed here.

\subsection{Undrained thermal deformations}
\label{sec:thermal:4.2}

\begin{figure}[htbp]
% Use the relevant command to insert your figure file.
% For example, with the graphicx package use
  \includegraphics[width=\columnwidth]{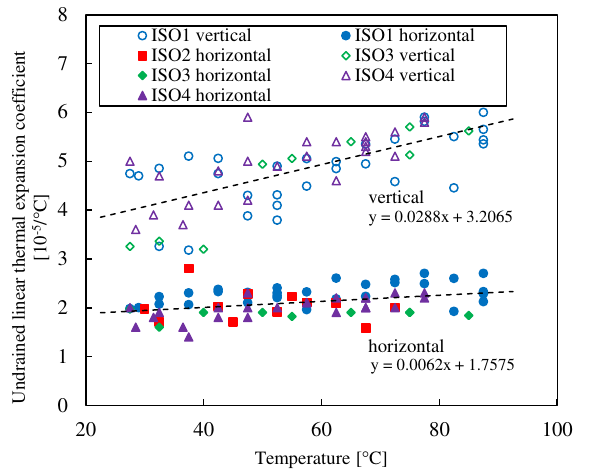}
% figure caption is below the figure
\caption{Linear undrained thermal expansion coefficient through heating and cooling, measured during the initial rapid temperature changes in each step test.}
\label{fig:thermal:alphauzh}       % Give a unique label
\end{figure}

During the first phase of each temperature step, the undrained thermal expansion coefficients were determined from the tangent of the strain-temperature curves. All four specimens showed a similar behaviour with axial strains about 2.3 times higher than radial ones, and an average value $\alpha_{u,z}=4.6 \times 10^{-5}  {^\circ \mathrm{C}}^{-1}$ (from 3 samples) and $\alpha_{u,h}=2.0 \times 10^{-5}  {^\circ \mathrm{C}}^{-1}$ (from 4 samples). By carrying out step tests at different temperatures, we can present the changes in thermal expansion coefficients with respect to temperature, as seen in Fig. \ref{fig:thermal:alphauzh}. 
\hz{The observed scatter of data can originate from the natural variability of properties, non-linearities of thermal expansion and measurement uncertainties due to the small deformations mobilized. For instance, the scatter in vertical direction corresponds to around 0.5 $\mu$m total vertical displacement on a 10 mm high specimen during a heating step of 5${^\circ \mathrm{C}}$.} 
Note that the values of the undrained thermal expansion coefficients are nearly one order of magnitude higher than the drained ones shown before. An increase in both vertical and horizontal thermal expansion coefficients with temperature was observed. A linear \hl{temperature} relationship \hl{of} $\alpha_{u,z}$ and $\alpha_{u,h}$ is \hl{found by a least square error fit} (Fig. \ref{fig:thermal:alphauzh}), and described with the following functions: 
\begin{equation}
\label{eq:thermal:}
\alpha_{u,z}=\left(0.028 T +3.256 \right) \times 10^{-5}  {^\circ \mathrm{C}}^{-1}
\end{equation}
\begin{equation}
\label{eq:thermal:}
\alpha_{u,h}=\left(0.006 T +1.758 \right) \times 10^{-5}  {^\circ \mathrm{C}}^{-1}
\end{equation}
with $T$ in ${^\circ \mathrm{C}}$ \hl{from 25 to 90 ${^\circ \mathrm{C}}$}.

This also provides an expression for the volumetric undrained thermal expansion coefficient \hl{$\alpha_{u}=2\alpha_{u,h}+\alpha_{u,z}$:}
\begin{equation}
\label{eq:thermal:alphau}
\alpha_{u}=\left(0.040 T +6.772 \right) \times 10^{-5}  {^\circ \mathrm{C}}^{-1}
\end{equation}

\begin{figure}[htbp]
% Use the relevant command to insert your figure file.
% For example, with the graphicx package use
  \includegraphics[width=\columnwidth]{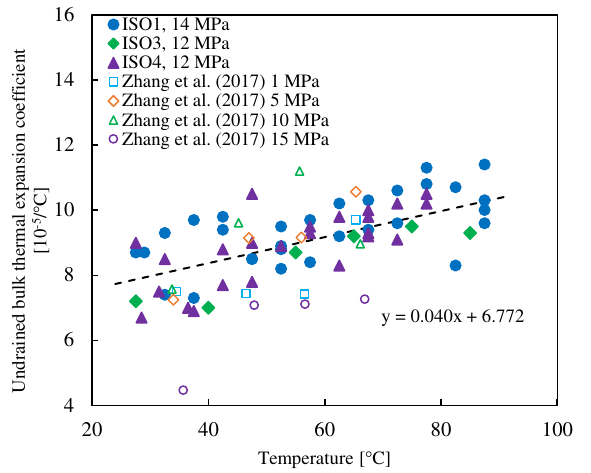}
% figure caption is below the figure
\caption{Bulk undrained thermal expansion coefficient through heating and cooling, measured during the initial rapid temperature changes in each step test. \hl{The results measured on specimens ISO 1, 3 and 4 are compared with very similar findings of \mbox{\cite{Zhang201746}}. Total isotropic stress levels are noted in the legend.}}
\label{fig:thermal:alphau}       % Give a unique label
\end{figure}

This linear relationship is depicted in Fig. \ref{fig:thermal:alphau} together with the measured bulk thermal expansion coefficients. Note that, due to the absence of radial measurements on ISO2, the volumetric response could not be displayed for this \hl{specimen}.

We could not detect any \hl{notable} difference between undrained thermal expansion coefficients during heating and cooling, showing that \hl{thermoplastic} deformations were not significant in undrained conditions. As shown \hl{in Fig. \ref{fig:thermal:drdefh},} the deformations of the solid matrix ($\alpha_{d,h}$,$\alpha_{d,z}$) contain some thermoplastic characteristics, but since $\alpha_{d}=2\alpha_{d,h}+\alpha_{d,z}$ is about ten times smaller than $\alpha_{u}$, \hl{their} effects on the latter are not significant. We can conclude, that undrained thermal strains are mainly governed by the reversible thermal volume changes of the fluid phase.

\hl{Comparable data on the undrained thermal bulk expansion coefficient of COx specimens have been presented by \mbox{\cite{Zhang201746}}. Their results under different confining stresses show values close to the ones obtained in this study, indicating also an increase of $\alpha_{u}$ with temperature (Fig. \ref{fig:thermal:alphau}).}
\cite{Mohajerani201211} investigated the undrained thermal behaviour of two COx samples in an isotropic cell. They did not record any specimen deformations, but they performed a back-calculation of $\alpha_{u}$ based on their measurements of induced pore pressures due to thermal pressurization. This theoretical $\alpha_{u}$ increased with temperature, with about $9.0 \times 10^{-5}  {^\circ \mathrm{C}}^{-1}$ \hl{(close to our findings)} at 30 ${^\circ \mathrm{C}}$ and 8 MPa effective stress, up to about $1.6 \times 10^{-4}  {^\circ \mathrm{C}}^{-1}$ \hl{(much higher than our findings)} at 80 ${^\circ \mathrm{C}}$ and 3.4 MPa effective stress.

\subsection{Strain response to pore pressure change}
\label{sec:thermal:H}
By changing the pore pressure under isostatic and isothermal conditions, one can measure the Biot modulus $H=-\Delta p_f / \Delta \varepsilon_v$. This was done, once the undrained equilibrium was reached in each three-stage step test, by opening the drainage system and bringing the thermally induced pore pressure back to initial value (see Fig. \ref{fig:thermal:expp} and \ref{fig:thermal:exstrain}). 

Note that the thermoplastic response observed above 40 ${^\circ \mathrm{C}}$ could not provide $H$ along heating paths, because $H$ is an elastic property. Consequently, above 40 ${^\circ \mathrm{C}}$, $H$ was determined from cooling paths \hl{only}.

\begin{figure}[htbp]
% Use the relevant command to insert your figure file.
% For example, with the graphicx package use
  \includegraphics[width=\columnwidth]{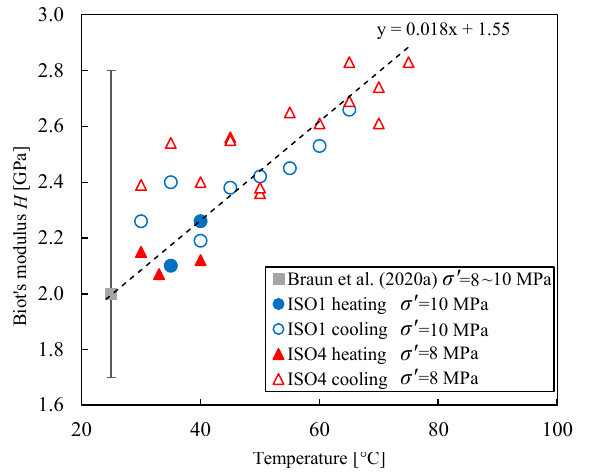}
% figure caption is below the figure
\caption{Measured modulus $H$ with respect to temperature. \hl{The square mark indicates the best-fit value from \mbox{\cite{Braun2020a}} at 25 ${^\circ \mathrm{C}}$ and 9 MPa effective stress, where the bars indicate the range of measured data. The trendline was obtained by a least square fit going through the best-fit value of \mbox{\cite{Braun2020a}}.}}
\label{fig:thermal:Htemp}       % Give a unique label
\end{figure}

As one can observe in Fig. \ref{fig:thermal:Htemp}, the measured moduli appear to increase with temperature. %, %which can be attributed to the thermal compaction observed before. 
\hz{The reported significant scatter of data is probably due to the natural variability of the material property, non-linear behaviour and measurement uncertainties on the relatively small deformations.}
The values at ambient temperature \hz{are in agreement} with the findings of \hl{our companion paper \mbox{\citep{Braun2020a}}, where} a mean value of $H$ around \hl{1.99 GPa (for isotropic effective stresses between 8 and 10 MPa at 25 ${^\circ \mathrm{C}}$) was presented. This evaluation was based on direct measurements and on a regression of various measured poroelastic properties. In the companion paper \mbox{\citep{Braun2020a}}, one can observe however some} variability of measured moduli at the same stress state, indicated in Fig. \ref{fig:thermal:Htemp}. The values measured in the present work lie almost all within this range, making it difficult to identify a clear temperature dependency.

\subsection{Thermally induced pore pressure}
\label{sec:thermal:thpress}

\begin{figure}[htbp]
% Use the relevant command to insert your figure file.
% For example, with the graphicx package use
  \includegraphics[width=\columnwidth]{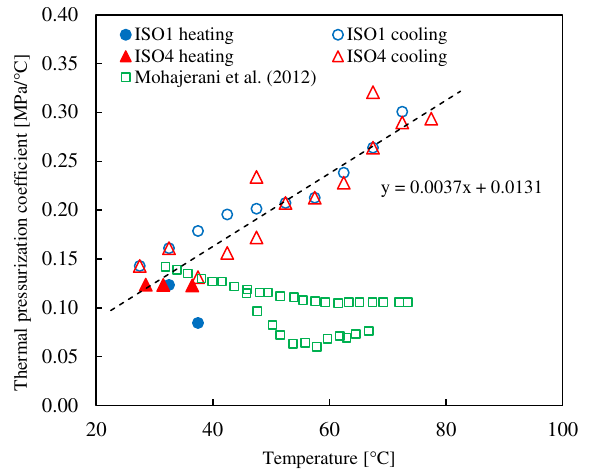}
% figure caption is below the figure
\caption{Corrected thermal pressurization coefficients obtained on samples ISO1 and ISO4 under heating up to 40 ${^\circ \mathrm{C}}$ and during cooling paths, compared with the findings of \cite{Mohajerani201211}.} 
\label{fig:thermal:Lambda}       % Give a unique label
\end{figure}

Using the calibration presented in Sec. \ref{sec:thermal:correction}, we corrected the thermal pressurization coefficients, measured from the temperature steps applied on samples ISO1 and ISO4. However, this correction method is based on the assumption of a thermo-elastic response, which was not the case during heating over 40 ${^\circ \mathrm{C}}$. Therefore, we show in Fig. \ref{fig:thermal:Lambda} the corrected results obtained along heating paths below 40 ${^\circ \mathrm{C}}$ and along cooling paths, during which the response is considered as \hl{thermo-elastic}. One can observe reasonable agreement between the data from heating below 40 ${^\circ \mathrm{C}}$ and \hl{from the} cooling phases. 

We can see that the parameter $\Lambda$ increases significantly with temperature, with values of around 0.12 MPa/${^\circ \mathrm{C}}$ at 25 ${^\circ \mathrm{C}}$ and up to around 0.30 MPa/${^\circ \mathrm{C}}$ at 80 ${^\circ \mathrm{C}}$. Note that the \hl{effective} stress conditions for these measurements remained nearly constant, close to the in-situ ones. The total stress was kept constant, while the thermally induced pore pressures were released after each step. 

\cite{Mohajerani201211} measured thermal pressurization coefficients of two COx \hl{specimens}, heated under initially close to in-situ stress ($\sigma'=$ 8 MPa). During the experiment, $\Lambda$ decreased slightly with temperature, with about 0.14 MPa/${^\circ \mathrm{C}}$ at 25 ${^\circ \mathrm{C}}$ down to between 0.10 and 0.05 MPa/${^\circ \mathrm{C}}$ above 70 ${^\circ \mathrm{C}}$. As seen in Fig. \ref{fig:thermal:Lambda}, their results for \hl{temperature below 40 $^{\circ}$C} correspond well to the findings of this study, however one can observe a large difference at higher temperatures. This difference will be discussed in more details in Sec. \ref{sec:thermal:thpress_analysis}.

\begin{figure}[htbp]
% Use the relevant command to insert your figure file.
% For example, with the graphicx package use
  \includegraphics[width=\columnwidth]{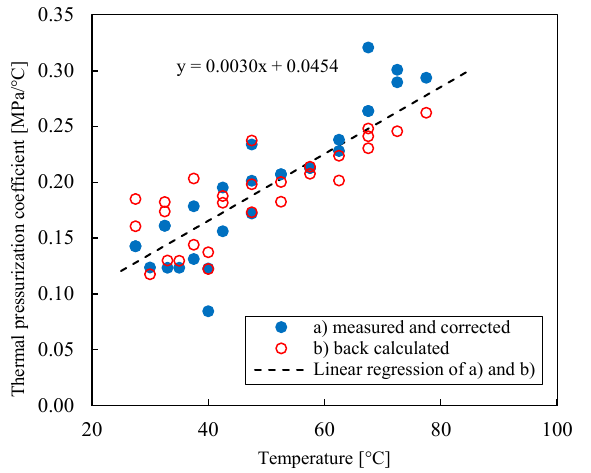}
% figure caption is below the figure
\caption{Comparison of the measured thermal pressurization coefficients (Fig. \ref{fig:thermal:Lambda}) with the back-calculated ones for each test using Eq. (\ref{eq:thermal:reluiso}).}
\label{fig:thermal:Lambdacomp}       % Give a unique label
\end{figure}

In parallel, we use Eq. (\ref{eq:thermal:reluiso}) to compute $\Lambda$ for each step test carried out within the thermo-elastic regime, based on the measurements of $\alpha_u$, $\alpha_d$ and $H$ from each respective test. As seen in Fig. \ref{fig:thermal:Lambdacomp}, these calculated values show good overall compatibility with the corrected measurements. Only the measured values for temperatures over 70 ${^\circ \mathrm{C}}$ appear to be slightly higher than the calculated ones. 
\hz{The observed scatter of the datapoints on the thermal pressurization coefficient $\Lambda$ and the differences between measured/corrected and back-calculated values are probably due to the combined error of the various thermo-poromechanical properties involved in the correction (Eq. (\ref{eq:thermal:corrLambda})) or in the back-calculation (Eq. (\ref{eq:thermal:reluiso})).}

\section{Discussion}
\label{sec:thermal:discuss}

\subsection{Thermal strains of the solid matrix}

Within a thermo-poromechanical framework, assuming micro-homogeneity, one can assume $\alpha_d=\alpha_{\phi}=\alpha_s$. The drained thermoelastic expansion coefficient of a specimen is therefore equal to the thermal volume changes of the solid matrix $\alpha_s$. In the case of COx claystone, for which we measured $\alpha_{d,z}=0.21 \times 10^{-5}  {^\circ \mathrm{C}}^{-1}$ (perpendicular to bedding) and $\alpha_{d,h}=0.51 \times 10^{-5}  {^\circ \mathrm{C}}^{-1}$ (parallel to bedding), the solid phase is composed of around 42 $\%$ clay minerals, 30 $\%$ calcite, 25 $\%$ quartz and 4 $\%$ feldspar \citep{Conil201861}.

\begin{table}
\caption{Thermal expansion coefficients of the bulk mineral composites and in directions perpendicular to the mineral layer ($\perp$) and parallel to the layer ($\parallel$), adapted from \cite{Belmokhtar201722}, \hz{compared with the properties of the COx claystone determined in this study}.}
\label{tab:1}
\begin{tabular}{llccc}
\hline\noalign{\smallskip}
Mineral          & Mass-\% &          \multicolumn{3}{c}{$\alpha$ $[10^{-5}{^\circ \mathrm{C}}^{-1} ]$ }       \\
\noalign{\smallskip}\cline{3-5}\noalign{\smallskip}
				&			&bulk			&linear ($\perp$)		&linear ($\parallel$)\\
\hline\noalign{\smallskip}
Clay 			 & 42      & 2.48            & 1.78 $^1$&0.35 $^1$\\
(muscovite)      &         &                 &        &  \\
Quartz           & 25      & 3.3 $^2$    &  \multicolumn{2}{c}{1.1}      \\
Calcite          & 30      & 1.4 $^3$	& \multicolumn{2}{c}{0.5} 		\\
Feldspar         & 4       & 1.1 $^3$	&\multicolumn{2}{c}{0.4} 		 \\
\hline\noalign{\smallskip}
\hz{COx claystone}   &  	      & \hz{1.23 $^4$}           & \hz{0.21 $^4$}  &\hz{0.51 $^4$}\\
\hline\noalign{\smallskip}
\multicolumn{5}{l}{$^1$\cite{McKinstry1965}, $^2$\cite{Palciauskas198228},}\\
\multicolumn{5}{l}{$^3$\cite{Fei199529}, \hz{$^4$this study}}
\end{tabular}
\end{table}
%\efloatseparator

\cite{McKinstry1965} measured the thermal expansion of clay minerals using a X-ray method. He provided mean values for muscovite, a mineral with the same molecular organisation as illite and smectite, of 1.78 $\times 10^{-5} {^\circ \mathrm{C}}^{-1}$ perpendicular to the layer and of 0.35 $\times 10^{-5} {^\circ \mathrm{C}}^{-1}$ parallel to the layer. \cite{Fei199529} gave values for the volumetric thermal expansion of calcite \hl{(1.4 $\times 10^{-5}  {^\circ \mathrm{C}}^{-1}$) and feldspar (1.1 $\times 10^{-5}  {^\circ \mathrm{C}}^{-1}$), and \mbox{\cite{Palciauskas198228}} for quartz (3.3 $\times 10^{-5}  {^\circ \mathrm{C}}^{-1}$) (see Tab. \ref{tab:1} for an overview).}

\hl{The thermal expansion coefficients of the mineral constituents of the COx claystone are compared in Tab. \ref{tab:1} with the measured drained thermal expansion coefficients. One can see that the drained thermal expansion coefficient parallel to bedding (0.51 $\times 10^{-5}  {^\circ \mathrm{C}}^{-1}$) lies within in the range of the values of elementary minerals (0.35 - 1.1 $\times 10^{-5}  {^\circ \mathrm{C}}^{-1}$). The reversible thermal expansion coefficient perpendicular to bedding (0.21 $\times 10^{-5}  {^\circ \mathrm{C}}^{-1}$) is however smaller than that of the mineral constituents (0.4 - 1.78 $\times 10^{-5}  {^\circ \mathrm{C}}^{-1}$).}
This fact, together with the contraction observed along heating paths perpendicular to bedding, shows that strains in this direction are not only dependent on the thermo-elastic behaviour of the minerals. Given the preferred sub-horizontal orientation of the clay platelets, we assume that adsorbed water molecules, either within the platelets or in the inter-platelets porosity, could have an influence on thermal strains perpendicular to bedding. \hl{Also the possible exchange of water between the solid phase and the bulk water through sorption is not captured by our model. The sorption potential has been found highest on smectites compared to other clay minerals, which are found in the form of 10-24 \% interstratified illite/smectite layers in the COx matrix. Note that the sorption potential can be also temperature dependent, which could be a responsible mechanism behind the observed nonlinear thermoplasticity (e.g. \mbox{\citealp{Brochard201712,Liu201845}}).} 

\subsection{Parametric study of THM couplings}

Looking at the thermoelastic expression of $\Lambda$ given in Eq. (\ref{eq:thermal:lambda}), one observes that the parameter $\Lambda$ is mainly depending on the difference between the thermal expansion of pore fluid $\alpha_f$ and of the solid matrix $\alpha_d$. The unjacketed modulus $K_s$, the fluid bulk modulus $K_f$, the Biot modulus $H$ and the porosity $\phi$ are also involved in the expression of $\Lambda$. 

\begin{figure}[htbp]
% Use the relevant command to insert your figure file.
% For example, with the graphicx package use
  \includegraphics[width=\columnwidth]{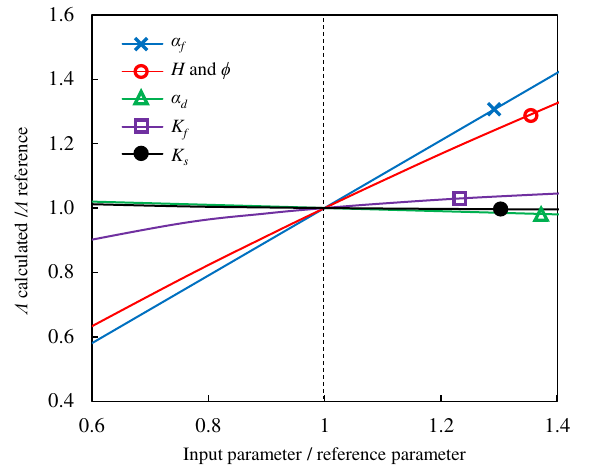}
% figure caption is below the figure
\caption{Variation of $\Lambda$ with respect to a separate variation of the input parameters listed in Tab. \ref{tab:thermal:Lambdasens}.}
\label{fig:thermal:Lambdasens}       % Give a unique label
\end{figure}

\begin{figure}[htbp]
% Use the relevant command to insert your figure file.
% For example, with the graphicx package use
  \includegraphics[width=\columnwidth]{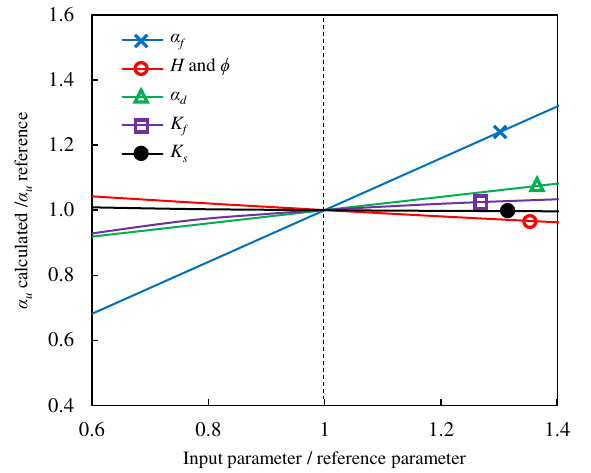}
% figure caption is below the figure
\caption{Variation of $\alpha_u$ with respect to a separate variation of the input parameters listed in Tab. \ref{tab:thermal:Lambdasens}.}
\label{fig:thermal:alphausens}       % Give a unique label
\end{figure}

A sensitivity analysis was carried out by calculating $\Lambda$ with Eq. (\ref{eq:thermal:lambda}), and $\alpha_u$ with Eq. (\ref{eq:thermal:reluiso}), using the previously considered reference parameters presented in Tab. \ref{tab:thermal:Lambdasens}. The influences of separate variation of each input parameter on $\Lambda$ and $\alpha_u$ are shown in Fig. \ref{fig:thermal:Lambdasens} and \ref{fig:thermal:alphausens}, respectively. One observes little sensitivity of both parameters with respect to $\alpha_d$, $K_s$ and $K_f$. Conversely, the thermal pressurization coefficient is strongly dependent on the parameters $\alpha_f$, $\phi$ and $H$, with the highest \hz{effect} caused by $\alpha_f$. We can also see that $\alpha_u$ is less sensitive to changes in $\phi$ and $H$, showing that $\alpha_f$ is the single main parameter governing the changes in $\alpha_u$.  

\begin{table}[htbp]
% table caption is above the table
\caption{Reference parameters used for the sensitivity analysis at 25 ${^\circ \mathrm{C}}$, \hl{$\sigma = 13$} MPa, $p_f = 4$ MPa. }
\label{tab:thermal:Lambdasens}       % Give a unique label
\begin{tabular}{lllp{3.5cm}}
\hline\noalign{\smallskip}
$\alpha_d$ 	& $[{^\circ \mathrm{C} ^{-1}}]$ &1.23$\times 10^{-5}$ 	&  measured\\
$\alpha_f$  & $[{^\circ \mathrm{C} ^{-1}}]$ &2.60$\times 10^{-4}$ 	& \cite{IAPWS-IF972008}\\
$K_s$     	& $[\text{GPa}]$  				&21.0      & \cite{Belmokhtar201787}     \\
$K_f$      	& $[\text{GPa}]$ 				&2.2      & \cite{IAPWS-IF972008}  \\
$\phi$    	& $[\%]$   		  				&18.0       & measured\\
$H$     	& $[\text{GPa}]$  				&\hl{1.99}     & \hl{\mbox{\cite{Braun2020a}}} \\
\noalign{\smallskip}\hline
\end{tabular}
\end{table}

\subsection{Anomalous thermal behaviour of pore fluid}
\label{sec:thermal:water}

\begin{figure}[htbp]
% Use the relevant command to insert your figure file.
% For example, with the graphicx package use
  \includegraphics[width=\columnwidth]{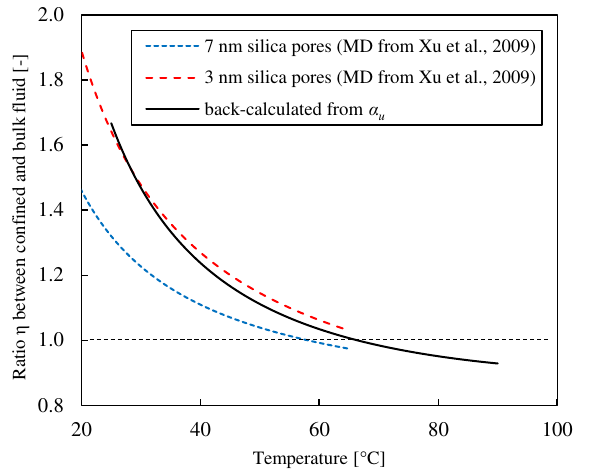}
% figure caption is below the figure
\caption{\hl{Ratio $\eta$ between the back-}calculated thermal expansion coefficient of the water confined in COx pores (Eq. (\ref{eq:thermal:af})) \hl{and} that of bulk water, compared to molecular dynamics (MD) simulations of \cite{Xu200950}.}
\label{fig:thermal:alphaf}       % Give a unique label
\end{figure}

\cite{Baldi198880} observed on low porosity clays, that the experimentally measured undrained response could only be reproduced by utilizing a thermal expansion coefficient of the pore fluid, higher than that of free water.
In the case of Opalinus clay, \cite{Monfared201173} observed the same phenomenon and back-calculated a thermal expansion coefficient of the pore fluid about two times higher than that of free water. \cite{Valenza200557} reported an anomalous thermal expansion of water, confined in cement pores, based on their thermopermeametry and beam bending experiments, while \cite{Ghabezloo200954} confirmed such findings with measurements from undrained heating tests on hardened cement paste.
In laboratory experiments on silica gel and silica glasses, respectively, \cite{Derjaguin199239} and \cite{Xu200950} could measure an elevated thermal expansion coefficient of the water contained in these small pores of nano-scale. \cite{Derjaguin199239} noted that, starting from 70 ${^\circ \mathrm{C}}$, the confined water lost its anomalous characteristics. \cite{Xu200950} were able to simulate their laboratory experiments up to 40 ${^\circ \mathrm{C}}$ by molecular dynamics calculations up to around 60 ${^\circ \mathrm{C}}$. Their extrapolation for higher temperatures, presented in Fig. \ref{fig:thermal:alphaf}, also show that the thermal expansion coefficient of confined water became equal to that of bulk water at around 70 ${^\circ \mathrm{C}}$. 

To analyse the thermal expansion property of COx pore fluid, we utilize the strong coupling between $\alpha_u$ and $\alpha_f$ (Fig. \ref{fig:thermal:alphausens}). This allows  us to carry out a back-analysis of $\alpha_f$, using the experimentally determined $\alpha_u$, by rewriting Eq. (\ref{eq:thermal:reluiso}) and (\ref{eq:thermal:lambda}):
\begin{equation}
\label{eq:thermal:af}
{\alpha _f} = \left( {\frac{1}{\phi } + H\left( {\frac{1}{{{K_f}}} - \frac{1}{{{K_s}}}} \right)} \right)\left( {{\alpha _u} - {\alpha _d}} \right) + {\alpha _d}
\end{equation}
We utilize Eq. (\ref{eq:thermal:af}) to calculate the thermal expansion coefficient of the pore fluid, under constant confining stress and pore pressure, as a function of  temperature. The parameters of Tab. \ref{tab:thermal:Lambdasens} are used here, taking into account the temperature and stress dependency of $K_f$ \citep{IAPWS-IF972008}. For $H$, we use the \hl{value at ambient temperature from Tab. \ref{tab:thermal:Lambdasens} and combine it with the observed temperature dependency (Fig. \ref{fig:thermal:Htemp}) to obtain ${H}(\sigma'=9 \, \mathrm{MPa},T) = 1.99+0.018( T-25  )$,} with $H$ in GPa and $T$ in ${^\circ \mathrm{C}}$.
Also the experimentally determined undrained thermal expansion coefficient (Fig. \ref{fig:thermal:alphau}) is inserted in Eq. (\ref{eq:thermal:af}). Doing so, we obtain a pore fluid thermal expansion coefficient $\alpha_f$, which is highly temperature dependent, and differs from the thermal expansion coefficient of free water. We introduce a factor $\eta = \alpha_f$(calulated)/$\alpha_f$(bulk water), which is shown in Fig. \ref{fig:thermal:alphaf}. For ambient temperatures, the pore fluid expands around 1.7 times more than bulk water. \hl{The ratio $\eta$} decreases with increasing temperature. For temperatures over 60 ${^\circ \mathrm{C}}$, the thermal expansion coefficient of the pore fluid is slightly lower than that of free water \hl{($\eta<1.0$).} A comparison with the values presented by \cite{Xu200950} shows a good \hl{agreement.}
\hl{This indicates a very peculiar behaviour of the water confined in the COx pores, to which also the sorption behaviour of the 10-24 \% mixed illite/smectite content could be contributing. A potential release of adsorbed water molecules from the solid phase to the fluid phase could be a reason for the increasing pore water volume expansion, which is not captured by the thermo-poro-elastic model.}

\subsection{Stress and temperature dependent thermal pressurization}
\label{sec:thermal:thpress_analysis}

\begin{figure*}[htbp]
% Use the relevant command to insert your figure file.
% For example, with the graphicx package use
  \includegraphics[width=\textwidth]{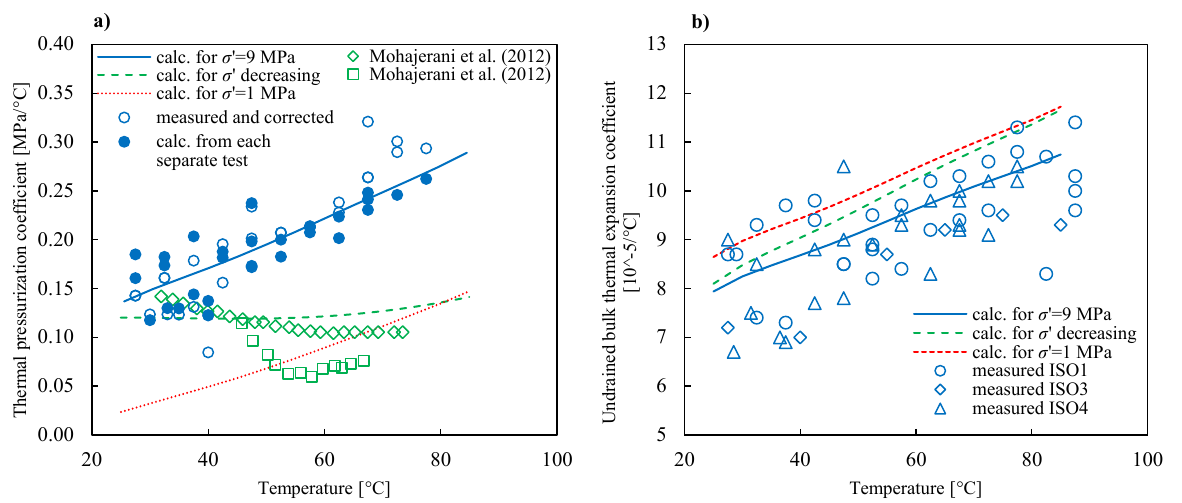}
% figure caption is below the figure
\caption{a) Back calculation of the thermal pressurization under different stress conditions, compared with the \hl{data at $\sigma'=8\sim10$ {MPa}} presented in this study and \hl{measured at decreasing effective stress} by \cite{Mohajerani201211}. b) Back-calculation of the undrained bulk thermal expansion coefficient under different stress conditions, compared with the measurements presented in this study. \hl{We used the values for $\alpha_f$ shown in Fig. \ref{fig:thermal:alphaf}.}}
\label{fig:thermal:Lambdacalc1}       % Give a unique label
\end{figure*}

To investigate the stress and temperature dependency of the thermal pressurization coefficient, we insert in Eq. (\ref{eq:thermal:lambda}) the previously back-calculated elevated thermal expansion coefficient of the pore fluid (Fig. \ref{fig:thermal:alphaf}). Again, the temperature and stress dependency for $K_f$ \citep{IAPWS-IF972008} was considered.
For the Biot modulus $H$, we use the following stress-dependent relationship from \cite{Braun2020a}:
\begin{equation}
{H}(\sigma', T=25^{\circ}\rm{C}) = \left[  4.070 \exp \left( { - 0.34 {\sigma '}}  \right) +0.311   \right] ^{-1}%
\label{eq:thermal:Hdmodel}
\end{equation}
\hl{with $H$ in GPa and ${\sigma '}$ in MPa. }
By assuming this stress dependency to be valid for all temperatures, we can add the temperature dependent part observed in this study (Fig. \ref{fig:thermal:Htemp}):
\begin{equation}
{H}(\sigma',T) = {H}(\sigma', T=25^{\circ}\mathrm{C})+ \kappa  \left( T - 25  \right)
\label{eq:thermal:Hdmodeltemp}
\end{equation}
\hl{with $\kappa= 0.018$, $H$ in GPa and $T$ in ${^\circ \mathrm{C}}$, obtained by a least square error fit.}
Other coefficients were taken from Tab. \ref{tab:thermal:Lambdasens}.
A significant increase of $\Lambda$ with temperature due to an increasing $\alpha_f$ can be \hl{noticed} in Fig. \ref{fig:thermal:Lambdacalc1}a \hl{(blue solid line).} We compare these values to our measurements (which were corrected for the effects of the drainage system, see Sec. \ref{sec:thermal:correction}) and to the calculated values from direct measurements using Eq. (\ref{eq:thermal:reluiso}), also presented in Fig. \ref{fig:thermal:Lambdacalc1}a. A good agreement between these values at temperatures up to 60 ${^\circ \mathrm{C}}$ can be observed. For higher temperatures however, the back calculation underestimates slightly the measured thermal pressurization.

Using the same \hl{parameters}, but changing the stress conditions according to the experiment carried out by \cite{Mohajerani201211}, we are able to reproduce a similar nearly constant parameter \hl{$\Lambda(\sigma',T)$}, plotted in Fig. \ref{fig:thermal:Lambdacalc1}a. \hl{Here we calculated $\Lambda$ through Eq. (\ref{eq:thermal:lambda}), first for $\Lambda_0(\sigma'_0=$ 8 $\mathrm{MPa},T_0=$ 25 $^{\circ} \mathrm{C})$. Then $\Lambda_{i+1}=\Lambda(\sigma'_{i}-\Lambda_i \mathrm{d}T,T_i+\mathrm{d}T)$ was computed subsequently with $\mathrm{d}T=$ 1 $^{\circ} \mathrm{C}$.}
The changed stress conditions, which include a continuously increasing pore pressure (according to $\Delta p_f=\Lambda \Delta  T$) reducing the effective stress, mainly affect the parameter $H$. A decreasing $H$ results here in a thermal pressurization coefficient, which remains nearly constant, evidencing a strong stress dependency of $\Lambda$. Another calculation for a constant, but very low effective stress, is presented in Fig. \ref{fig:thermal:Lambdacalc1}a. Due to constant effective stress, we see again an increasing thermal pressurization coefficient, with however much lower values. Due to the decreasing effective stress, the material becomes more compliant. The increase of the fluid thermal expansion coefficient with temperature is counterbalanced by the higher mechanical deformation of the pore space.  

The effect of stress on $\alpha_u$ was investigated by calculating $\alpha_u$ using Eq. (\ref{eq:thermal:relu}) and (\ref{eq:thermal:reluiso}), as shown in Fig. \ref{fig:thermal:Lambdacalc1}b. Unsurprisingly, since the measured mean values of $\alpha_u$ were used as input parameters for the model, we find a good agreement for the calculation at \hl{$\sigma'=9$} MPa. More interestingly, we observe only a small stress dependency of $\alpha_u$ at $\sigma'=1$ MPa, which lies within the range of variability of measured values. The plot for a decreasing effective stress represents, as also for the thermal pressurization coefficients in Fig. \ref{fig:thermal:Lambdacalc1}a, some transition between the limiting curves at \hl{9} and 1 MPa effective stress. 

\section{Conclusions}
\label{sec:thermal:conclusion} 
 
Time saving testing protocols for thermal laboratory experiments, together with an optimized experimental device, allowed \hl{us} to efficiently determine the thermal properties of the COx claystone on four \hl{specimens}. A modified testing apparatus with a reduced drainage length permitted relatively fast specimen saturation and drainage times. The application of the protocols of \cite{Braun2019} provided multiple material parameters in each experiment, rendering the laboratory study more time efficient. Strain gages were employed for measuring precise specimen deformations, where it was crucial to account for thermally induced measurement errors, especially due to the relatively small thermal strain response of the tested material.
Reversible drained thermal strains, parallel to the bedding plane, were detected during heating and cooling up to 90 $^\circ \mathrm{C}$. In \hz{the} perpendicular direction, we observed a thermal contraction upon heating, which appeared to be irreversible. During cooling, the material contracted in both directions, with strains parallel to bedding about 2.4 times higher than perpendicular to bedding. A relatively small reversible volumetric drained thermal expansion coefficient $\alpha_d = 1.23\times 10^{-5}{^\circ \mathrm{C} ^{-1}}$ was found.
The undrained thermal expansion coefficient during heating and cooling under constant stress conditions was determined, and found to increase with temperature. A good  compatibility with the transversely isotropic drained and undrained thermal expansion coefficients and the Biot modulus $H$, which were simultaneously measured for each test, gives confidence in the measured properties.    
By conducting a back analysis, we showed that for reproducing the undrained thermal expansion coefficient and the thermal pressurization coefficient in a thermo-poro-elastic framework, one has to consider a thermal expansion coefficient of the pore water greater than that of bulk water. Together with the measured Biot modulus $H$ and the measured drained thermal expansion coefficient $\alpha_d$, the thermal pressurization coefficient $\Lambda$ was calculated. The theoretical values reflect well experimental results, and \hz{can} reproduce a thermal pressurization coefficient that increases significantly with temperature and decreases with decreasing effective stress. By decreasing the effective stress during undrained heating, the material stiffness decreases, expressed through $H$, which results in a nearly constant thermal pressurization coefficient. One can also note that undrained characteristics ($\Lambda$, $\alpha_{u,i}$) were insensitive to $\alpha_{d,i}$, due to the relatively small values of $\alpha_{d,i}$. This emphasizes the decisive role of pore fluid in the undrained thermal behaviour of the COx claystone. The anomalous thermal fluid expansion, the thermoplastic behaviour of the solid matrix \hl{and the coupling with stress dependent poroelastic properties} should be considered in engineering models, when analysing the response of the formation to thermal loading.

\section*{Conflict of interest}

The authors declare that there are no known conflicts of interest associated with this publication.

\bibliographystyle{spbasic}      % basic style, author-year citations
\bibliography{bib_article}   % name your BibTeX data base

% Non-BibTeX users please use
%\begin{thebibliography}{}
%
% and use \bibitem to create references. Consult the Instructions
% for authors for reference list style.
%
%\bibitem{RefJ}
%% Format for Journal Reference
%Author, Article title, Journal, Volume, page numbers (year)
%% Format for books
%\bibitem{RefB}
%Author, Book title, page numbers. Publisher, place (year)
%% etc
%\end{thebibliography}

\end{document}